\documentclass[10pt,letter,dvipsnames]{article}
\usepackage{amsmath,amssymb}
\usepackage{geometry}
\usepackage{color}
\usepackage{graphicx}
\usepackage{algorithm}
\usepackage{algpseudocode}
\usepackage{soul}
\usepackage{url}
\usepackage{hyperref}
\usepackage{setspace}
\usepackage{tikz}
\usepackage{xcolor}
\usetikzlibrary{positioning}

\usepackage{multicol}
\newcommand{\CC}{\mathcal{C}}

\newcommand{\LC}{\mathcal{L}}
\DeclareMathOperator*{\argmin}{arg\,min}

\algrenewcommand\algorithmicrequire{\textbf{Inputs:}}
\algrenewcommand\algorithmicensure{\textbf{Output:}}

\title{Low-Complexity Ordered Reliability Direct Error Pattern Testing (ORDEPT) Decoding with Likelihood Thresholding}

\author{Reza Hadavian and Dmitri Truhachev\footnote{The authors are with the Dept. of Electr. and Comp. Engineering, Dalhousie University, Canada,  \{reza.hadavian,dmitry\}@dal.ca}}
\date{}

\makeatletter
\renewcommand\@bibitem[1]{\item\if@filesw \immediate\write\@auxout{\string\bibcite{#1}{\the\value{\@listctr}}}\fi\ignorespaces}
\makeatother

\begin{document}

\maketitle
\thispagestyle{empty}

\section{Introduction} \label{sec:Intro}
The design of new low-complexity and latency soft decoding algorithms for short block codes is receiving increased research interest due to developments in high-speed optical and massive wireless communications. Fiber-optical communication modules that reach speeds of terabit per second on a single wavelength require low-complexity, near-error-free error-correction coding, typically constructed by concatenation of short codes~\cite{400ZR,wang2022investigation}. Massive machine-to-machine (M2M) communications, remote industrial automation, Internet-of-Things (IoT), virtual reality, telehealth, and many other applications within the realm of 5G and 6G mobile communications necessitate reliable ultra low-latency short packet communications. Therefore, the traditional ordered statistic decoding (OSD) and the industry-standard Chase decoding \cite{chase1972class} have recently been complemented by a variety of new algorithms such as soft versions of guessing random additive noise decoding (GRAND) \cite{solomon2020soft,duffy2022ordered}, direct error-pattern testing (DEPT) and ORDEPT \cite{truhachev2020efficient,hadavian2025ordered}, and guessing codeword decoding (GCD) \cite{ma2024guessing,zheng2024universal}. Typically, these algorithms test error patterns in the order of their approximate likelihood until one or several candidate codewords are found and the best candidate is selected. The complexity of such pattern-based search is largely determined by the number of queries and the operations associated with testing of each query. Lowering testing complexity is crucial for making the hardware implementation of such algorithms competitive.

In this work we propose a low-complexity ORDEPT decoding with likelihood-based termination. The decoding operates as follows: 1) Instead of testing complete error patterns in the order of their likelihood, the algorithm tests partial error patterns which are missing one error position. The final position is found on-the-fly by a combinational logic function, reducing the overall number of queries dramatically. 2) In addition, once a candidate codeword is found, its log-likelihood difference to the received sequence is compared to a preset threshold and the decoding decision is instantly made in case the likelihood deviation is below the threshold. 

We demonstrate that while keeping the same block error rate (BLER) performance, the proposed algorithm's latency and complexity is multiple times smaller than that of all other state-of-the art competitors including the Chase II, ORBGRAND, GCD and the very recent ORDEPT with Soft-Output GRAND termination~\cite{rapp2025sogrand} which necessitates several multiplications in each query processing. The results for decoding of  Bose–Chaudhuri–Hocquenghem (BCH) $(256,239)$ code serving as a component code in 400ZR+ \cite{400ZR,wang2022investigation}, BCH(32,21) codes, and (128,116) Polar code are demonstrated in Section~\ref{sec:simulation}. We can observe that the proposed algorithm's BLER performance is at par with the competitors', while the numbers of required operations (computational complexity) are orders of magnitude smaller.

\section{System Model} \label{sec:system}

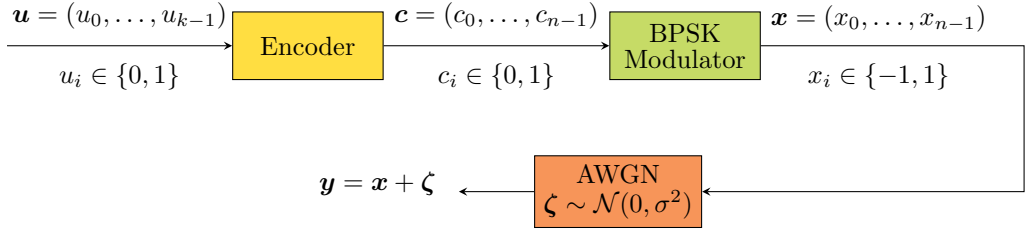
\begin{figure}[htbp]
\centering
\begin{tikzpicture}
\node [draw,
    fill=Goldenrod,
    minimum width=2cm,
    minimum height=0.9cm,
] (encoder) at (3,0) {$\mathrm{Encoder}$};
 
\node [draw,
    fill=SpringGreen, 
    minimum width=2cm, 
    minimum height=0.9cm,
    right=3cm of encoder
] (modulator) {\shortstack{BPSK \\ Modulator}};
 
\node [draw,
    fill=Peach, 
    minimum width=2cm, 
    minimum height=0.9cm, 
    below right= 1cm and 2cm of encoder
]  (AWGN) {\shortstack{AWGN \\ $\boldsymbol{\zeta}\sim\mathcal{N}(0,\sigma^2)$}};
 
\draw[-stealth] (-1,0) -- (encoder.west)
    node[midway,above,yshift=3pt]{$\boldsymbol{u}=(u_0,\ldots,u_{k-1})$}
    node[midway, below, yshift=-3pt] {$u_i\in \{0,1\}$};

\draw[-stealth] (encoder.east) -- (modulator.west) 
    node[midway,above,yshift=3pt]{$\boldsymbol{c}=(c_0,\ldots,c_{n-1})$}
     node[midway, below, yshift=-3pt] {$c_i\in \{0,1\}$}; 


    \draw[-stealth] (modulator.east) -- ++ (3.5,0) |- (AWGN.east)
        node[midway,above,yshift=57pt,xshift=-55pt]{$\boldsymbol{x}=(x_0,\ldots,x_{n-1})$}
        node[midway, below, yshift=52pt,xshift=-55] {$x_i\in \{-1,1\}$};

        \draw[-stealth] (AWGN.west) -- ++(-1cm,0) 
            node[midway,above,,yshift=-5pt,xshift=-45pt]{$\boldsymbol{y}=\boldsymbol{x}+\boldsymbol{\zeta}$};

\end{tikzpicture}
    \caption{Block diagram of system model.}
    \label{Fig:systemmodel}
\end{figure}

Consider a binary information vector of length $k$, which is encoded using a binary linear $(n,k)$ block code $\CC$. The encoding procedure produces an $n$-bit codeword $\boldsymbol{c}=({c}_0,{c}_1,\ldots,{c}_{n-1})$, where ${c}_i \in \{0,1\}$ for $i\in\{0,\ldots,n-1\}$. The codeword $\boldsymbol{c}$ is subsequently modulated using binary phase-shift keying (BPSK). Each element $c_i$ of the codeword is mapped to $x_i\in \{-1,1\}$ according to the relation $x_i=1-2c_i$. The resulting modulated signal $\boldsymbol{x}=({x}_0,{x}_1,\ldots,{x}_{n-1})$ is transmitted over a real-valued additive white Gaussian noise (AWGN) channel. The channel output is expressed as
\begin{equation} \label{eq:channel_output}
\boldsymbol{y}=\boldsymbol{x}+\boldsymbol{\zeta},
\end{equation}
where $\boldsymbol{\zeta}$ is a vector of independent and identically distributed (i.i.d.) Gaussian random variables with zero mean and variance $\sigma^2$. The system model is depicted in Fig \ref{Fig:systemmodel}. Using (\ref{eq:channel_output}), the hard-decision received sequence $\boldsymbol{w}$ is obtained by computing $\boldsymbol{w}=\frac{1}{2}(\mathrm{sign}(\boldsymbol{y})+1)$. The noise effect, represented as ${\boldsymbol{z}}$, is defined as the difference between the transmitted codeword $\boldsymbol{c}$ and the hard-decision sequence $\boldsymbol{w}$, i.e., ${\boldsymbol{z}}=\boldsymbol{c}\oplus\boldsymbol{w}$.

We denote the absolute values of the received sequence bit LLRs as $|\boldsymbol{l}|=(|l_0|, |l_1|,\ldots, |l_{n-1}|)$, where $l_i=2y_i\sigma^{-2}$ for $i=0,1,\ldots,n-1$. These values can be sorted in ascending order using a permutation $\boldsymbol\pi=(\pi_0, \ldots, \pi_{n-1})$.
Depending on how the LLR vector and the ordering permutation $\boldsymbol{\pi}$ are utilized, various universal soft-decoding techniques can be identified, which aim to recover $\boldsymbol{x}$ from $\boldsymbol{y}$.

\section{Decoding Algorithms}

Consider the following decoding algorithms, recently proposed in literature.

\subsection{ORBGRAND}
   
    Operates on a list of test error patterns $\hat{\boldsymbol{e}}^{(1)},\ldots,\hat{\boldsymbol{e}}^{(q)}, \ldots$, where $\hat{\boldsymbol{e}}^{(q)}=(\hat{e}_0^{(q)},\ldots,\hat{e}_{n-1}^{(q)})$, $\hat{e}_i^{(q)} \in \{0,1\}$, are ordered from most likely to list likely. The algorithm checks if $(\boldsymbol{w} \oplus \tilde{\boldsymbol{e}}^{(Q)}) { \bf H}^\textrm{T} =  \boldsymbol{0}$. The first codeword for which this equality holds will be output. In the list ORBGRAND \cite{abbas2022list}, the decoder continues the search to find $C_{\mathrm{max}}$ codewords where $C_{\mathrm{max}}$ is maximum number of found codewords. Then, the most likely codeword is output. 


\subsection{ORDEPT}

The ORDEPT algorithm operates on a list of PEPs instead of full error patterns. A PEP, represented as $\boldsymbol{e}$, is a binary vector where each element $e_i$ indicates a potential error at the $i$-th position by being set to $1$. However, the final error position — which is associated with the lowest likelihood and thus the highest index — remains $0$, along with all other components of $\boldsymbol{e}$. The key idea is to determine whether a given PEP can be completed by appending an index for the last error position, resulting in a valid codeword. Applying a PEP represented by $\boldsymbol{e}$, which contains $N-1$ ones at the positions $e_{j_0}, e_{j_1}, \ldots, e_{j_{N-2}}$, to the hard-decision received sequence $\boldsymbol{w}$ results in the computation of a partial syndrome vector as follows:
\begin{equation} \label{eq:syndrome_prime}
\tilde{\boldsymbol{s}}=(\boldsymbol{w} \oplus {\boldsymbol{e}}) { \bf H}^\textrm{T}=\boldsymbol{s} \oplus \boldsymbol{h}{j_0}\oplus \boldsymbol{h}{j_1} \oplus \cdots \oplus \boldsymbol{h}_{j{N-2}},
\end{equation}
where $\boldsymbol{h}_{j_i}$ denotes the $j_i$-th column of the code's parity-check matrix $\bf{H}$ for $i = 0, \ldots, N-2$, and $\boldsymbol{s}$ is the $(n-k)$-bit syndrome vector defined by $\boldsymbol{s} = \boldsymbol{w} {\bf H}^\textrm{T}$.

This partial syndrome $\tilde{\boldsymbol{s}}$ is subsequently used to determine the final error position, denoted as $j_*$, which completes the PEP $\boldsymbol{e}$ so that the full error pattern $\tilde{\boldsymbol{e}}$ yields a valid codeword. Here, $\tilde{\boldsymbol{e}}$ represents the completed error pattern such that $\tilde{e}_j = e_j$ for all $j \neq j_*$, and $\tilde{e}_{j_*} = 1$. Identifying $j_*$ means finding a column in ${\bf H}$ for which $\tilde{\boldsymbol{s}} = \boldsymbol{h}_{j_*}$. This association is performed using a function $f(\tilde{\boldsymbol{s}})$ that maps the column vector $\boldsymbol{h}_{j} = \tilde{\boldsymbol{s}}$ of $\bf{H}$ to its corresponding index $j_*$:
\begin{equation} \label{eq:analytic_function}
f(\tilde{\boldsymbol{s}})=\begin{cases}
j_* & \mathrm{if}\quad \tilde{\boldsymbol{s}}=\boldsymbol{h}_{j_*}\\
-1 & \mathrm{otherwise.}
\end{cases}
\end{equation}

A result of $f(\tilde{\boldsymbol{s}}) = -1$ indicates that no column index $j_*$ satisfies $\boldsymbol{h}_{j_*} = \tilde{\boldsymbol{s}}$. When $j_*$ is successfully located and $f(\tilde{\boldsymbol{s}}) \neq -1$, the corrected codeword is given by
\begin{equation} \label{eq:error_correction}
\tilde{\boldsymbol{c}}=\boldsymbol{w}\oplus\tilde{\boldsymbol{e}}.
\end{equation} 

The performance of the proposed algorithm can be enhanced if it continues to search for additional candidate codewords after finding the first one. In such a scenario, the LLR summation of the completed error patterns, that is denoted via $\epsilon(\tilde{\boldsymbol{c}})$ are calculated, and the candidate with the smallest LLR summation is selected as the decoded codeword.

\subsection{ORDEPT with SOGRAND-based thresholding}
For each decoding, SOGRAND provides an a-posteriori probability that a codeword is correct or the decoding is not in the list. Denoting the a-posteriori probability of each guessed word via

\begin{equation}
    p(\boldsymbol{y}^{(i)}|\boldsymbol{l})=\prod_{j=0}^{n-1} 
    \left\{
    \begin{array}{ll}
        \displaystyle \frac{1}{1+\mathrm{exp}(-l_j)}, & \quad y_j^i=0 \\[1em]

        \displaystyle \frac{1}{1+\mathrm{exp}(l_j)}, & \quad y_j^i=1
    \end{array}
    \right.,
\end{equation}

during the guessing procedure, the algorithm accumulates the probabilities of the guesses $\boldsymbol{y}^{(i)}$ from

\begin{equation} \label{eq:pnoise}
    p_\mathrm{noise} = \sum_{i=1}^{q}p(\boldsymbol{y}^{(i)}|\boldsymbol{l}),
\end{equation}

where $q$ is the number of guesses. Let $\mathcal{L}\subset\mathcal{C}$ be a list of the codewords that the decoder has found until the $q$-th guess. Following the results in \cite{rapp2025sogrand}, the probability that the correct codeword is not in the list is estimated as

\begin{equation} \label{eq:notinlist}
    p(\boldsymbol{\boldsymbol{c}\notin\mathcal{L|\boldsymbol{l}}}) = \frac{(1-(p_{\mathrm{noise}}+\sum\limits_{\hat{\boldsymbol{c}}\in\mathcal{L}}p(\hat{\boldsymbol{c}}|l)))2^{k-n}}{\sum\limits_{\hat{\boldsymbol{c}}\in\mathcal{L}}p(\hat{\boldsymbol{c}}|l)+(p_{\mathrm{noise}}+\sum\limits_{\hat{\boldsymbol{c}}\in\mathcal{L}}p(\hat{\boldsymbol{c}}|l)))2^{k-n}}.
\end{equation}


The approach considered in \cite{rapp2025sogrand} employs a termination criterion based on SOGRAND, which estimates the probability that the correct codeword has been identified after each new codeword is added to the list. Decoding stops if either $p(\boldsymbol{\boldsymbol{c}\notin\mathcal{L|\boldsymbol{l}}})\leq\theta$, where $\theta$ is a predefined threshold $\theta$, or the decoder finds $C_{\mathrm{max}}$ codewords. The most likely codeword, $\boldsymbol{c}_\mathrm{best}=\argmin_{\hat{\boldsymbol{c}}\in\LC} p(\hat{\boldsymbol{c}}|\boldsymbol{l})$ is then output. This leads to a decoder with a dynamic list size that adjusts the amount of guesswork according to the specifics of each decoding instance.

\section{Proposed ORDEPT with likelihood-based thresholding}
In the proposed algorithm, sum of the absolute values of LLRs (analog weight of the respective error pattern) $\epsilon(\boldsymbol{c})$ is employed to estimate the likelihood of each found codeword. On this basis, each time a new codeword $\tilde{\boldsymbol{c}}$ entered the decoder's list, its analog weight is computed and compared against a predefined fixed threshold denoted by $\epsilon_T$. In case $\epsilon(\hat{\boldsymbol{c}})<\epsilon_T$, decoding is halted and $\tilde{\boldsymbol{c}}$ is the output. Otherwise, the decoder continues further search until 1) $q=Q_{\mathrm{max}}$ 2) $|C|=C_{\mathrm{max}}$. In this case, the most likely codeword in terms of analog weight is the output. The steps of proposed algorithm are detailed in Algorithm \ref{alg:ORDEPT}. The proposed algorithm reduces complexity compared to ORDEPT with SOGRAND-based termination by limiting additional operations to instances when a new codeword is found, whereas SOGRAND incurs extra operations for every query. 

\begin{algorithm}[H]
\caption{ORDEPT with likelihood-based thresholding}\label{alg:ORDEPT}
\begin{algorithmic}[1]
\Require {LLR vector $\boldsymbol{l}$, code $\mathcal{C}$, the maximum number of error pattern queries $Q_{\mathrm{max}}$}, PEP list, LLR threshold $\epsilon_T$, $C_{\mathrm{max}}$ number of candidate codewords to be found.
\Ensure {Decoded codeword $\boldsymbol{c}_\mathrm{best}$.}
\If{$\boldsymbol{w}\in\mathcal{C}$}
\textbf{return} $\boldsymbol{c}_\mathrm{best}=\boldsymbol{w}$
\Else \hspace{3mm}Find the positions $\pi_0,\pi_1,\pi_2,\ldots$ of the smallest absolute LLR values $|\boldsymbol{l}|$.
\State $q=0$  \Comment{Initialize the index of the current PEP query.} 
\State $C=0$  \Comment{Initialize the number of codewords found by the algorithm.} 
\State Calculate the syndrome $\boldsymbol{s} = \boldsymbol{w} {\bf H}^\textrm{T}$.
\While{$C<C_{\mathrm{max}}$ \textbf{and}  $q<Q_{\mathrm{max}}$} Generate or retrieve the next PEP $\boldsymbol{e}^{(q)}$ from the PEP list and apply $\boldsymbol{\pi}$ to acquire ${\boldsymbol{e}}^{(q)}$. Obtain $\tilde{\boldsymbol{s}}$.
    \If{The current PEP can be completed} complete it with $j^*$. Compute $\tilde{\boldsymbol{c}}$ and $\epsilon(\tilde{\boldsymbol{c}})$.
         \If{$\epsilon(\tilde{\boldsymbol{c}})<\epsilon_T$}
         \textbf{return} $\boldsymbol{c}_{\mathrm{best}}=\tilde{\boldsymbol{c}}$.
    \EndIf
    \State Store $\tilde{\boldsymbol{c}}$ and $\epsilon(\tilde{c})$ $C=C+1$, \quad $q=q+1$
    \Else
    $\hspace{2mm} q=q+1$ 
    \EndIf
\EndWhile
\State \textbf{return} $\boldsymbol{c}_\mathrm{best}=\argmin_{\tilde{\boldsymbol{c}}\in\mathcal{L}} \epsilon(\tilde{\boldsymbol{c}})$.
\EndIf
\end{algorithmic}
\end{algorithm}

For the optimized parameter selection of the proposed algorithm, we first set $\epsilon_T=0$ and choose the maximum number of candidate codewords $C_{\mathrm{max}}$ such that 

\begin{equation}
    C_{\mathrm{max}}^*= \mathrm{min} \big\{C_{\mathrm{max}}\in \mathbb{N}:\forall E_b/N_0\quad \mathrm{BLER}_{\mathrm{ORDEPT}}<\mathrm{BLER}_\mathrm{ref}\big\}.
\end{equation}

Subsequently, $C_\mathrm{max}=C_\mathrm{max}^*$ and $\epsilon_T$ is adjusted per $E_b/N_0$ such that

\begin{equation}
    \epsilon_{\mathrm{T}}^*= \mathrm{max} \big\{\epsilon_{\mathrm{max}}\in \mathbb{R}:\quad \mathrm{BLER}_{\mathrm{ORDEPT}}<\mathrm{BLER}_\mathrm{ref} \big\}.
\end{equation}
where $\mathrm{BLER}_\mathrm{ref}$ is the reference BER from another algorithm such as ORBGRANd, for example.

\section{Simulation Results} \label{sec:simulation}

This Section presents simulation results that illustrate the performance achieved by the proposed ORDEPT with likelihood-based thresholding in terms of BLER and decoding complexity (computed via the average number of queries and the number of arithmetic operation per query) for decoding of representative BCH and Polar codes compared to the state-of-the-art approaches. All simulations are conducted assuming transmission over the AWGN channel with BPSK signaling.
The 1-line ORBGRAND is used as test pattern (in ORBGRAND) and PEP (in ORDEPT) generator for query-based decoders. 

For complexity comparison, we consider the average number of queries and the number of arithmetic operations. 
ORDEPT with SOGRAND-based termination needs $n$ operations to calculate $p(\boldsymbol{0}|\boldsymbol{r})$, that is, the probability of the all-zero test pattern. Then, each time a PEP $\boldsymbol{e}^i$ is generated, its probability $p(\boldsymbol{e}^i|\boldsymbol{r})$ needs to be computed and added to $p_\mathrm{noise}$ as in \eqref{eq:pnoise}. The former requires $\omega(\boldsymbol{e}^i)$ operations where $\omega(\boldsymbol{e}^i)$ denotes the Hamming weight of $\boldsymbol{e^i}$ and the latter needs $q$ operations where $q$ is the number of queries. In the case of finding a candidate codeword, two additional operations are required to compute $p(\boldsymbol{\boldsymbol{c}\notin\mathcal{L|\boldsymbol{l}}})$ in \eqref{eq:notinlist}. One for adjusting the PEP probability with respect to the last error position, and the other for the addition. For ORDEPT with likelihood-based termination, only in the case of finding a candidate codeword, it requires computation of $\epsilon(\tilde{\boldsymbol{c}})$ which is obtained by summing the LLR quantities of nonzero positions in the completed PEP $\tilde{\boldsymbol{e}}$, which are usually sparse.

\begin{figure}[h]
    \setlength{\unitlength}{1mm}
    \begin{picture}(0,75)(0,0)
    \put(1,-5){\includegraphics[scale=0.085]{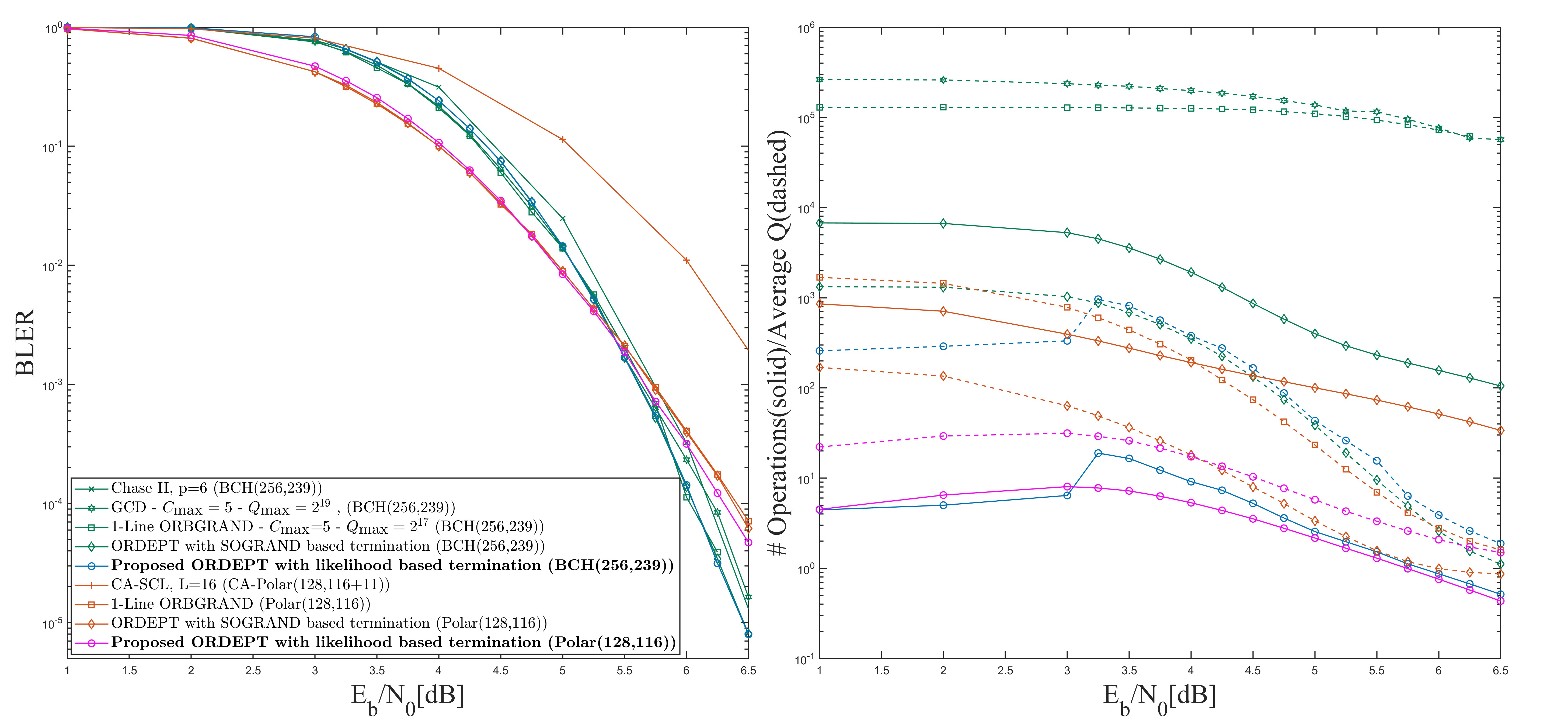}}
    \end{picture}
    \caption{Performance evaluation of ORDEPT in comparison with existing decoders for $\mathrm{BCH}(256,239)$.}
    \label{Fig:BLERBCHPolar}
\end{figure}

Fig.~\ref{Fig:BLERBCHPolar} shows the BLER performance (left panel) and complexity (right panel) of the proposed ORDEPT with likelihood-based thresholding in comparison with state-of-the-art decoders for $\mathrm{BCH}(256,239)$ and $\mathrm{Polar}(128,116)$. 


\begin{figure}[H]
    \setlength{\unitlength}{1mm}
    \begin{picture}(0,75)(0,0)
    \put(-2,-5){\includegraphics[scale=0.102]{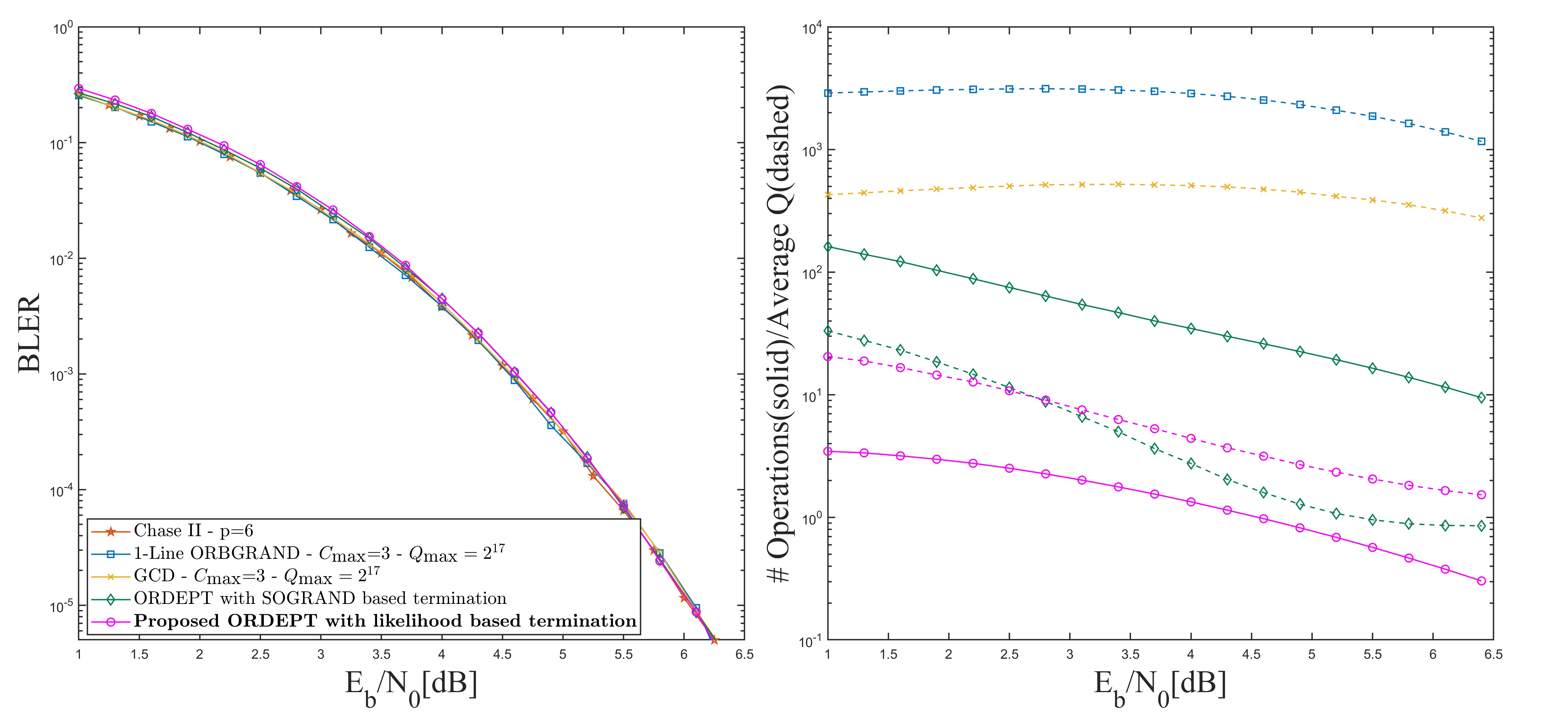}}
    \end{picture}
    \caption{Performance evaluation of ORDEPT in comparison with existing decoders for $\mathrm{BCH}(32,21)$.}
    \label{Fig:BLERBCH3221}
\end{figure}

In Fig.~\ref{Fig:BLERBCH3221} we compare the performance and complexity of the proposed ORDEPT with likelihood-based thresholding in comparison with state-of-the-art decoders for $\mathrm{BCH}(32,21)$. 

As it is seen from the presented results, the proposed algorithm maintains the same BLER performance as GCD and ORDEPT with SOGRAND-based thresholding while decreasing number of required operations by more than an order of magnitude compared to GCD and ORDEPT with SOGRAND thresholding. 

\end{document}